\documentclass[twocolumn,tighten,twocolappendix]{aastex631}
\usepackage{graphicx}
\usepackage{url}
\usepackage{color}
\usepackage{multirow}
\usepackage{ragged2e}
\usepackage{hyperref}
\usepackage{url}
\usepackage{amsmath} 
\usepackage{booktabs}
\usepackage{comment}
\usepackage{afterpage}
\usepackage{mathtools}
\usepackage{tabularx}
\usepackage{bold-extra}
\usepackage{xspace}
\usepackage{relsize}
\usepackage{newunicodechar}
\usepackage{lipsum}

\usepackage[encapsulated]{CJK}
\usepackage{ucs}
\usepackage[utf8x]{inputenc}

\DeclareGraphicsExtensions{.pdf,.png,.jpeg}

\definecolor{ForestGreen}{rgb}{0.133,0.545,0.133}

\def\rp{r_\gamma}
\def\m87{M87*}
\def\sgra{\mbox{Sgr A*}\xspace}

\begin{document}

\title{Photon Rings and Shadow Size for General Axi-Symmetric and Stationary Integrable spacetimes}

\author[0009-0006-0070-1888]{Kiana Salehi}
\affiliation{Perimeter Institute for Theoretical Physics, 31 Caroline Street North, Waterloo, ON, N2L 2Y5, Canada}
\affiliation{Department of Physics and Astronomy, University of Waterloo, 200 University Avenue West, Waterloo, ON, N2L 3G1, Canada}
\affiliation{Waterloo Centre for Astrophysics, University of Waterloo, Waterloo, ON N2L 3G1 Canada}

\author[0000-0002-3351-760X]{Avery E. Broderick}
\affiliation{Perimeter Institute for Theoretical Physics, 31 Caroline Street North, Waterloo, ON, N2L 2Y5, Canada}
\affiliation{Department of Physics and Astronomy, University of Waterloo, 200 University Avenue West, Waterloo, ON, N2L 3G1, Canada}
\affiliation{Waterloo Centre for Astrophysics, University of Waterloo, Waterloo, ON N2L 3G1 Canada}

\author[0000-0002-3586-6424]{Boris Georgiev}
\affiliation{Perimeter Institute for Theoretical Physics, 31 Caroline Street North, Waterloo, ON, N2L 2Y5, Canada}
\affiliation{Department of Physics and Astronomy, University of Waterloo, 200 University Avenue West, Waterloo, ON, N2L 3G1, Canada}
\affiliation{Waterloo Centre for Astrophysics, University of Waterloo, Waterloo, ON N2L 3G1 Canada}

\begin{abstract}

There are now multiple direct probes of the region near black hole horizons, including direct imaging with the Event Horizon Telescope (EHT).  As a result, it is now of considerable interest to identify what aspects of the underlying spacetime are constrained by these observations.  For this purpose, we present a new formulation of an existing broad class of integrable, axisymmetric, stationary spinning black hole spacetimes, specified by four free radial functions, that makes manifest which functions are responsible for setting the location and morphology of the event horizon and ergosphere.  We explore the size of the black hole shadow and high-order photon rings for polar observers, approximately appropriate for the EHT observations of \m87, finding analogous expressions to those for general spherical spacetimes.  Of particular interest, we find that these are independent of the properties of the ergosphere, but does directly probe on the free function that defines the event horizon.  Based on these, we extend the nonperturbative, nonparametric characterization of the gravitational implications of various near-horizon measurements to spinning spacetimes.  Finally, we demonstrate this characterization for a handful of explicit alternative spacetimes.
\end{abstract}


\section{Introduction} \label{sec:intro}
Astrophysical black holes have had a number of new windows opened upon the astrophysical and gravitational processes that play out near their event horizons.  Most explicitly, this has taken the form of the horizon-resolving images of the supermassive black holes at the center of M87 and the Milky Way, \m87 and \sgra, respectively, obtained by the Event Horizon Telescope \citep[EHT;][]{M87_PaperI,SgpaperI}.  The detection of gravitational waves by LIGO originating from the merger of stellar mass black holes probe similar spatial regions, though from black holes between 4 and 7 orders of magnitude less massive \citep{LIGOQNM}.  Both sets of observations are fundamentally impacted by the dynamics of massless fields near the event horizon, and are therefore potentially sensitive probes of the structure of black hole spacetimes in the strong-gravity regime.

EHT images of \m87 and \sgra reveal a bright annulus of emission, surrounding the black hole ``shadow'', the collection of photon trajectories that intersect the photon sphere, and thus terminate on the event horizon, and are therefore dark.  Its boundary is defined by the asymptote of an infinite series of lensed images of the emission region, the so-called ``photon rings''.  
The shadow is a purely gravitational feature.  The high-order photon rings are asymptotically insensitive to the astrophysics of the emission region.  Therefore, these features hold particular interest as a potential probe of gravity.

General relativity (GR) makes clear predictions for both EHT and LIGO observations.  For the EHT targets, in which the accretion flow is gravitationally inconsequential, the no hair theorems require that the spacetime is fully described by the Kerr metric \citep{Newman1965,Kerr1963}.  A similar expectation holds for the late-time ringdown portion of LIGO merger events \citep{LIGOQNM}.  Searches for deviations from this expectation have taken the form of explicitly constraining perturbations to the Kerr spacetime, arising either due to an alternative gravity theory or as part of an expansion \citep{SgpaperVI,Prashant2021,Johannsen2016,Broderick2014}.  The conclusions are, therefore, inseparably tied to the applicability of the underlying perturbation.

Weak-field gravity tests are typically characterized in a general, but perturbative, framework, the Parametrized Post-Newtonian (PPN) expansion \citep{Will2014}.  This has the significant benefit of being independent of the underlying details of the alternative gravity theory, and typically interpretable in terms of the foundational assumptions of GR \citep{Will2014}.  Developing a similar expansion in the strong field regime is difficult due to the inherently nonlinear nature of gravity near black hole event horizons: all PPN terms are of order unity at the horizon, and thus perturbative expansions are not useful.
\footnote{Test-specific expansions have been developed, e.g., the parametrized post-Einsteinian expansion of gravitational waveforms to study the implications of gravitational wave observations of merginig black holes in a model agnostic framework \citep{cornish2011}.}

To address the strong-gravity regime, a number of Kerr-like metrics have been constructed. A broad class of axisymmetric, stationary spacetimes is presented in \citet{Johannsen2013}, defined by four (nearly) arbitrary radial functions.  This class of metrics has the convenient feature that it is integrable, simplifying the construction of null geodesics, and therefore images.  As we demonstrate in \autoref{app:altmetrics}, the \citet{Johannsen2013} proposal encompasses or is equivalent to a large number of other classes of metrics found in the literature \citep[e.g., those of][]{Visser2023, CarsonYagi2020} and contains a number of specific alternative spacetimes \citep[e.g.][etc.]{Bardeen1973}. However, previous authors have sought to once again constrain the arbitrary radial functions via perturbative expansions that introduce strong assumptions about the nature and behavior of deviations from Kerr.

In \citet{SphericalShadows23}, we introduced a proposal for a nonperturbative and nonparametric scheme for characterizing the gravitational implications of EHT shadow measurements, future photon ring measurements, and LIGO QNM measurements.  The proposed scheme had the considerable virtue of not requiring extraordinary assumptions regarding the magnitude and form of deviations from general relativity, and this sense presented a means to recast near-horizon observations as direct measurements of the spacetime structure at the location of the stationary circular photon orbit, $\rp$.  It also rendered explicit what features were, and were not, constrained by existing near-horizon observations.  It was, however, only presented for spherically symmetric spacetimes, which is unlikely to be appropriate for EHT targets or the LIGO merger products probed by the ringdown.

Here we expand on the proposal in \citet{SphericalShadows23}, extending it to the class of spacetimes from \citet{Johannsen2013}.  We focus attention on polar observers, appropriate for \m87 which is viewed from approximately $17^\circ$ \citep{M87_PaperV}.  
Despite the increased complexity of this class of spacetimes, we maintain the two key features of our previous work:
\begin{enumerate}
    \item It remains nonparametric to retain full generality and avoid complications from strongly correlated parameters.
    \item It remains a nonperturbative characterization for better applicability in the highly-nonlinear near-horizon regime, for which there is no natural sense of ``small''.
\end{enumerate}
We find generalizations of the expressions for the shadow size and Lyapunov exponents presented in \citet{SphericalShadows23} that apply in the spinning spacetime case, and similar benefits accrue: the problem of collecting implications for multiple Kerr-alternatives is reduced to computing appropriate metric components at $\rp$, and multiple fundamentally distinct observations may be naturally combined.

We defer on the practical complication that the shadow size itself is not measurable; only the visible surrounding plasmas is observed.  We note that the calibration with general relativistic simulations performed in \citet{M87_PaperVI} and \citet{SgpaperVI} complicates their interpretation.  Direct methods for extracting higher-order images, i.e., the $n=1$ photon ring, have been proposed and applied to \m87, from which the shadow boundary may be inferred \citep{Rings,Themaging,Spin,Johnson2020}. However, we set aside the measurement difficulties and continue to focus on the narrow question: what would we learn from such a measurement?

The paper is structured in the following manner: In \autoref{sec:polar}, we compute the null polar trajectories in spacetimes that are spherically symmetric and static, making contact with \citet{SphericalShadows23} and building intuition relevant for the more general case. In \autoref{sec:metric}, we extend our attention to the class of integrable, axisymmetric and stationary spacetimes in \citet{Johannsen2013}, recast in a form that explicitly separates the free functions that define the event horizon and ergosphere.  Subsequently, we determine the radius of the shadow size in \autoref{sec:shadow}
and the high-order photon rings in \autoref{sec:rings}. In \autoref{sec:obs}, we utilize the EHT measurements to constrain our modified metric. We present our conclusions in \autoref{sec:conclusions}.  Appropriate forms of well-known alternative metrics are collected in \autoref{app:altmetrics}.

\section{The Polar Observer in General Spherically Symmetric Spacetimes}
\label{sec:polar}

Spherically symmetry renders the problem of computing orbits planar.  As a consequence, it is typical to compute orbits in the equatorial plane.  However, because the situation of practical interest here are polar orbits about axisymmetric spacetimes, this simplification is no longer possible.  Before discussing the general case, we begin by demonstrating the equivalence of such a calculation in the spherically symmetric case, introduce the relevant constants of motion, and provide a explicit example of their use.

As in \citet{SphericalShadows23}, we begin with a general spherically symmetric, stationary metric, expressed in areal coordinates,
\begin{equation}
    ds^2=-N^2(r) dt^2+\frac{B^2(r)}{N^2(r)} dr^2+ r^2 d \Omega,
    \label{eq:spherical_symmetric_metric}
\end{equation}
where $N(r)$ and $B(r)$ are arbitrary functions of $r$.

\subsection{Shadow Size}

\subsubsection{Equatorial Photon Orbits}
Photon orbits that lie within the equatorial plane are fully described by two constants of the motion, associated with the timelike and azimuthal killing fields,
\begin{equation}
    e = p_t = -N^2(r) \dot{t}
    ~~\text{and}~~
    L_z = p_\phi = r^2 \dot{\phi},
\label{eq:consts}
\end{equation}
where $\dot{f}\equiv df/d\lambda$ is differentiation with respect to an affine parameterization and $p_\mu$ is the covariant photon four-momentum.  These two constants correspond to the energy and angular momentum, respectively, and have their normal non-relativistic meanings at infinity.  Specifically, $e=h\nu_{\rm obs}$ is the photon energy measured by an observer at infinity, and $\ell \equiv L_z/e$ is the specific angular momentum about the polar axis.  We note further, that for photon orbits that reach infinity, $\ell$ is also equal to the impact parameter.

In terms of these constants, the equations of motion of massless particles reduce to,
\begin{equation}
    \dot{r}^2 =
    \frac{1}{B^2(r)}
    \left[ 1 - \frac{\ell^2}{r^2} N^2(r) \right]
    ~~\text{and}~~
    \dot{\phi} = \frac{\ell}{r^2},
\label{eq:equitorial_eoms}
\end{equation}
where without loss of generality we have set $e=1$.
The radius of the unstable circular photon orbit, $\rp$, may now be found by asserting $\dot{r}=0$ and $\ddot{r}=0$, giving the result
\begin{equation}
    \rp = \frac{N(\rp)}{N'(\rp)},
\end{equation}
with angular momentum
\begin{equation}
    \ell = \frac{1}{N'(\rp)},
\end{equation}
as found in \citet{SphericalShadows23}.  For photons that graze $\rp$, i.e., have an innermost radial turning point at this radius, the impact parameter is $R=\ell$, and we have the normal result for the radius of the ``shadow'', i.e., the dark region comprised of trajectories that would originate on the event horizon.

\subsubsection{Polar Photon Orbits}
We now repeat the previous computation, but in a language explicitly relevant for polar orbits.  Again, we require two constants of the motion, the first of which will be the energy, $e$ (we we again set to unity).  The second is now given by the Carter constant,
\begin{equation}
    Q = p_\theta^2 + L_z^2\cot^2\theta,
    \label{eq:carter_sphsymm}
\end{equation}
which for orbits with $L_z=0$ reduces to $Q=p_\theta^2$.  We again define $q\equiv \sqrt{Q}$, and the equations of motion for the polar trajectories (with $L_z=0$) are
\begin{equation}
    \dot{r}^2 = \frac{1}{B^2(r)} \left[1-\frac{q^2}{r^2} N^2(r) \right]
    ~~\text{and}~~
    \dot{\theta} = \frac{q}{r^2},
\label{eq:polar_eoms}
\end{equation}
which apart from replacing one constant for another ($q$ for $\ell$) is identical to \autoref{eq:equitorial_eoms}.  It immediately follows that $\rp=N(\rp)/N'(\rp)$ and $q=1/N'(\rp)$.  As before, for polar observers, $q$ may be identified with the impact parameter for photon orbits that reach observers at infinity.

The shadow is surrounded by an infinite series of higher-order images, i.e., the ``photon rings''.  Each subsequent order ring is associated with an additional half-orbit made by the null geodesic before it heads towards a faraway observer. As each ring, defined this way, is situated at a unique position in the image, no position can produce more than one null geodesic or image.\footnote{Higher-order rings arising from a finite emission region, need not execute exactly half-integer numbers of orbits, and therefore, can intersect.  This degeneracy becomes asymptotically suppressed with photon-ring order.} 
These features are due to the strong gravitational lensing within the spacetime and serve as an excellent tool with which to investigate general relativity.

\subsection{High-order Photon Rings}

We may go further, and compute the Lyapunov exponent of radial orbits near $\rp$ and their relationship to the high-order photon rings.  In the equatorial plane, the argument proceeds as detailed in Appendix D of \citet{SphericalShadows23}; here we derive that result using the polar orbits.

We expand the radius about $r=\rp+\delta r$ and perturbatively expand \autoref{eq:polar_eoms}, to obtain,
\begin{equation}
    \dot{\delta r}^2 =
    - \frac{q^2}{2B^2(r)} \left[
    \frac{N^2(r)}{r^2}
    \right]''_{\rp} \delta r^2,
    \label{eq:spherical_symmetric_dr}
\end{equation}
As with the shadow size, this is identical to the equatorial calculation apart from substituting $q$ for $\ell$.  The corresponding expression for the Lyapunov exponent for radial orbits is
\begin{equation}
    \gamma = \pi \frac{N^{3/2}(r_\gamma)}{N'(r_\gamma)} 
    \left[ -\frac{N''(r_\gamma)}{B^2(r_\gamma)} \right]^{1/2},
\end{equation}
as found in \citet{SphericalShadows23}.

This illustrates the general procedure we will follow in the remainder of this paper, and the role that the Carter constant will play, analogous to that of angular momentum in the spherically context.

\section{A General Non-Parametric non purtubative axisymmetric spacetime}
\label{sec:metric}

The Kerr metric is an exact solution to the vacuum Einstein field equations, and describes spinning black hole spacetimes.  It is the sole uncharged stationary, axisymmetric, asymptotically flat, nonpathological solution to the the Einstein equations, and therefore is believed to be applicable to astrophysical black holes.\footnote{While the Kerr-Newman metric presents more general class of spacetimes, describing charged black holes, the abundance of free charges in astrophysical environments precludes this possibility in practice.}

By virtue of its two explicit Killing symmetries (associated with stationarity and axisymmetry), by Noether's theorem test particles have two integrals of motion: energy and angular momentum around the axis of rotation.  Less obvious is the presence of a third integral of motion, the Carter constant \citep{Carter1968}.  In addition to a fourth constant, particle rest mass, test particle orbits in Kerr are fully integrable.

In contrast, deviations from Kerr need not admit fully integrable geodesics.  Nevertheless, their obvious utility has motivated the definition of a broad family of arbitrary stationary, axisymmetric metrics that have four constants of motion by  \citet{Johannsen2013}.  By construction, this set of metrics possesses a modified Carter constant, allowing us to avoid concerns about the integrability of the spacetime and simplifying the examination the path of photons around the central mass. These metrics introduces four arbitrary real functions of radius \citep[called $A_1(r)$, $A_2(r)$, $A_5(r)$, and $f(r)$ in][]{Johannsen2013} that we will not expand upon and will leave in their general form. Consequently, we extend the proposal a nonperturbative, nonparametric characterization of shadow size and related measurements in the spacetime domain in \citet{SphericalShadows23} to a broad class of spinning spacetimes, explicitly demonstrating the nature and effectiveness of shadow size-based constraints.

We begin with a manifestly stationary and axisymmetric metric, expressed in terms of three four arbitrary functions of radius, $N(r)$, $B(r)$, $F(r)$, and $f(r)$,
\begin{equation}
    \begin{split}
    ds^2 = &
    -\frac{\tilde{\Sigma} N^2}{(r-F a^2\sin^2\theta)^2} (dt-a\sin^2\theta d\phi)^2\\
    +& \frac{\tilde{\Sigma}\sin^2\theta}{(r-F a^2\sin^2\theta)^2} (aF dt - r d\phi)^2\\
    +  & \frac{\tilde{\Sigma}}{r^2} \frac{B^2}{N^2} dr^2
    + \tilde{\Sigma} d\theta^2,
    \label{eq:metric}
   \end{split}
\end{equation}
where
\begin{equation}
\begin{aligned}
    \tilde{\Sigma} &= r^2 + a^2\cos^2\theta + f(r)\\
    \Delta &= r^2-2Mr+a^2,\\
\end{aligned}
\end{equation}
and the parameters $M$ and $a$ are parameters related to the mass and spin\footnote{$M$ and $a$ are exactly the ADM mass and spin when $N(r)$, $B(r)$, $F(r)$, and $f(r)$ take on appropriate asymptotic behavior at large $r$ (see \autoref{app:asympADM}).}.
This metric is identical to that presented in Equation~51 of \citet{Johannsen2013} (see \autoref{app:metric_comp}).

In \autoref{app:altmetrics}, we further demonstrate that it is either identical to or encompasses a number of other proposed families of stationary axisymmetric metrics from the literature. 
These include those proposed by \citet{Baines2023} and \citet{CarsonYagi2020} (after applying the observational constraints discussed near Equation 50 of \citealt{Johannsen2013}), and the regular black hole spacetime in \citet{AzregAinou2014}.  For each of these metric families, the mapping of their free functions to $N(r)$, $B(r)$, $F(r)$, $f(r)$ is summarized at the bottom of \autoref{tab:altmetrics}.  That many authors have independently arrived at equivalent families suggests that this form is a general representation of regular, stationary and axisymmetric spacetimes.  This is not the case, however, and examples exist that admit integrable null geodesics but are expressible in the form \autoref{eq:metric}, e.g., the \citet{Rasheed1995}.  Nevertheless, a large number of commonly discussed alternative spacetimes may be placed in the form of \autoref{eq:metric}, as detailed in \autoref{tab:altmetrics} and \autoref{app:altmetrics}

Expressing the \citet{Johannsen2013} family of metrics in this fashion has some significant conceptual benefits.  
\begin{enumerate}
\item The event horizon, defined by where $g_{rr}$ diverges, occurs at the largest radius for which $N(r)=0$.  That is, $N(r)$ completely describes the location and structure of the event horizon; that an event horizon exists reduces to the requirement that $N(r)$ has at least one root.  Furthermore, the event horizon is necessarily at fixed $r$, i.e., within our particular choice of coordinates (which reduce to Boyer-Lindquist in Kerr), the horizon appears spherically symmetric.
\item The ergosphere, defined by where $g_{tt}$ vanishes, is set by
\begin{equation}
N^2(r) - a^2 F^2(r) \sin^2\theta = 0,
\label{eq:ergo}
\end{equation}
and thus its structure determined by $F^2(r)$ once $N^2(r)$ is specified, with immediate consequences for the existence and structure of the ergosphere.  When $a=0$, \autoref{eq:ergo} is never satisfied outside of the event horizon, and there is no ergosphere.  Similarly, when $F(r)=0$, there cannot be an ergosphere.  Thus, within this family of metrics, the spin and $F(r)$ must be non-zero for energy extraction via the Penrose process and related mechanisms  \citep{PenroseProcess}.  There is also a strong constraint on the poloidal shape of the ergosphere.  Because $N^2(r)/F^2(r)$ is a function of $r$ alone, the shape of the boundary of each ergosphere depends only the magnitude of the black hole spin (as it must be) and symmetric about the equatorial plane.
\item When $a=0$, $N(r)$ and $B(r)$ take on the same meaning as they did in \autoref{eq:spherical_symmetric_metric}.  For polar observers, we shall see they take on the same importance for the shadow and photon ring sizes.
\end{enumerate}

The first two constants of integration of photon trajectories are given by their usual expressions, $e \equiv p_t$ (which we set to unity henceforth) and $L_z=p_\phi = \ell$.  The third constant of motion is the modified Carter constant, which for photons is by construction,
\begin{equation}
    Q = p_\theta^2 + \frac{(\ell-a \sin^2\theta)^2}{\sin^2\theta} - (\ell-a)^2.
\end{equation}
For polar zero-angular-momentum observers, to which we will restrict ourselves, the Carter constant reduces to $Q=p_\theta^2-a^2$, which, as in the spherically symmetric case, we will again express in terms of $q\equiv\sqrt{Q}$.

\section{Polar Shadow Size}
\label{sec:shadow}
As seen by a distant observer located along the polar axis, the image of the event horizon (or the photon orbit) fills a circle.  The location of photon trajectories on a screen at large $r$ asymptotes to,
\begin{equation}
    R = \sqrt{q^2 + (\ell-a)^2 - (\ell-a\sin^2\theta)^2/\sin^2\theta},
\end{equation}
\citep[see][]{Bardeen1973}.  We will consider only orbits with $\ell=0$, which correspond to photon trajectories that are normal to a non-rotating screen at infinity, and a polar observer ($\theta=0$), for which $R$ reduces to
\begin{equation}
    R = \sqrt{q^2+a^2},
\end{equation}
reducing the problem of determining the shadow size to identifying the Carter constant of those photons that just graze the photon sphere, the generalization of the unstable circular photon orbit from the spherically symmetric case.

The equation of motion of the photons near the black hole is given in Equation~35 of \citet{Johannsen2013}, which expressed in terms of our metric functions becomes,
\begin{equation}
    \dot{r}^2 = 
    \frac{r^4}{\tilde{\Sigma}^2}
    \frac{1}{B^2(r)} \left[
    1 - \frac{q^2+a^2}{r^2} N^2(r)
    \right],
    \label{eq:j13_polar_eom}
\end{equation}
which differs from the spherically symmetric case only by a prefactor.  On the unstable photon sphere, we again require $\dot{r}=0$ and $\ddot{r}=0$.  The first is satisfied outside the event horizon when $q_\gamma^2+a^2 = \rp^2 / N^2(\rp)$.  The second requires
\begin{equation}
    \rp = N(\rp)/N'(\rp),
    \label{eq:rp}
\end{equation}
(see \autoref{app:rp}).
As the name photon sphere implies, in these coordinates, these orbits all lie at constant radius.  
Identifying the shadow radius with $R=\sqrt{q_\gamma^2+a^2}$, upon inserting the expression for $\rp$ gives
\begin{equation}
    R = \frac{1}{N'(\rp)}.
    \label{eq:R}
\end{equation}

Note that these expressions are {\em identical} to those for the spherically symmetric spacetime, and thus $N(r)$ plays an exactly analogous role in the family of integrable stationary axisymmetric spacetimes for polar observers.  As found for spherically symmetric spacetimes, the the size of the shadow solely relies upon the value of $N'(r)$ at $r_\gamma$.  

Therefore, shadow measurements do not directly constrain the size $N(r)$.  Furthermore, shadow measurements provide no information regarding the other arbitrary functions of the general metric, $B(r)$, $F(r)$, or $f(r)$.  Consequently, for example, shadow measurements alone are insensitive to the location and properties of the ergosphere, a point to which we will return in what follows.

\section{Multiple photon rings}
\label{sec:rings}
The shadow is surrounded by an infinite series of photon rings, composed of higher-order images. These rings are created by the null geodesic's execution of an additional half-orbit before moving towards a faraway observer. 

The radii of these features are geometric and depend on the strong gravitational lensing and asymptotically insensitive to the astrophysics of the emission region, making them an interesting tool for studying general relativity at the vicinity of the photon orbit.

Higher order photon rings lie exponentially closer to the shadow boundary, with their structure heavily impacted by the dynamics of orbits nearby $\rp$.  To examine the radii of the photon rings, we will make the following assumptions regarding their propagation:
\begin{enumerate}
\item Each photon ring has an innermost radial turning point at some radius outside of the photon sphere, $\rp+\delta r_0$.
\item This location is sufficiently close to the photon sphere that we may perturbatively expand the photon equation of motion about $\rp$.
\item Upon propagating some distance of order $M$ away from $\rp$, the photon will stream to the observer at infinity.
\end{enumerate}

The second provides an immediate equation of motion for $\delta r$, obtained from expanding \autoref{eq:j13_polar_eom} about $\rp$,
\begin{equation}
    \dot{\delta r}^2
    =
    \left.
    -\frac{r^4}{\tilde{\Sigma}^2}
    \frac{(q_\gamma^2+a^2)}{2 B^2(r)} \left[
    \frac{N^2(r)}{r^2} 
    \right]''
    \right|_{\rp} \delta r^2.
\label{eq:dr}
\end{equation}
Again, apart from prefactors, this is very similar to the expression found in the spherically symmetric case (see \autoref{eq:spherical_symmetric_dr}).  Importantly, it is proportional to $R^2 [N^2(r)/r^2]''$.  

Unlike the spherically symmetric case, this now depends on the polar angle via $\tilde{\Sigma}$.  However, the angular equation of motion is

\begin{equation}
    \dot{\theta} = p^\theta = \frac{1}{\tilde{\Sigma}}\sqrt{R^2-a^2 \sin^2 \theta},  
\end{equation}
where we have used the fact that $R=\sqrt{q_\gamma^2+a^2}$ is the radius of the shadow.  Therefore, the factors of $\tilde{\Sigma}$ cancel identically, resulting in an equation for $\delta r$ that is separable in $\theta$:
\begin{equation}
    \frac{d\delta r}{d\theta}
    =
    G(\theta)
    \frac{N^{3/2}(\rp)}{N'(\rp)}
    \left[-\frac{N''(\rp)}{B^2(\rp)}\right]^{1/2}
    \delta r,
\label{eq:ddrdth}
\end{equation}
where
\begin{equation}
    G(\theta) \equiv \frac{R}{\sqrt{R^2-a^2\sin^2\theta}}.
\end{equation}
We note that $G(\theta)$ is independent of the four metric perturbations, apart from the specification of $R$.  Therefore, we may integrate \autoref{eq:ddrdth} immediately to obtain that the growth in orbital radius from the $n$ to $n+1$ half orbit is
\begin{equation}
    \ln\left(\frac{\delta r_{n+1}}{\delta r_n}\right)
    =
    \frac{N^{3/2}(\rp)}{N'(\rp)}
    \left[-\frac{N''(\rp)}{B^2(\rp)}\right]^{1/2}
    \int_0^\pi d\theta G(\theta).
\end{equation}
Expressing this in terms of the orbital radius at the innermost radial turning point, the orbital radius after $n$ half-orbits is
\begin{equation}
    \delta r_n = \delta r_0 e^{\gamma n}
\end{equation}
in which the Lyapunov exponent is given by
\begin{equation}
    \gamma \equiv 
    \frac{N^{3/2}(\rp)}{N'(\rp)}
    \left[-\frac{N''(\rp)}{B^2(\rp)}\right]^{1/2}
    2 K[(a/R)^2]
    \label{eq:gamma}
\end{equation}
where $K(k)$ is the complete elliptic integral of the first kind \citep[see, e.g.][]{Abramowitz1972}.  Identically, when $a=0$, we have $K[(a/R)^2]=\pi/2$ and $\gamma$ reduces to the expression in Equation~22 of \citet{SphericalShadows23}.

The requirement that there is an inner turning point at $r=\rp+\delta r_0$ modifies the Carter constant associated with the photon ring trajectories.  Because $\dot{r}=0$ on the photon sphere, the lowest order at which the Carter constant changes is second order, i.e.,
\begin{equation}
    q^2+a^2 = q_\gamma^2 + a^2 + \frac{1}{2} \left[\frac{r^2}{N^2(r)}\right]''_{\rp} \delta r_0^2,
\end{equation}
Employing the third assumption, that upon reaching some radius ($\rp+\delta r_{\rm max}$) the photon streams toward infinity, we may estimate the radii of the photon rings generated by $n$ half orbits (i.e., $\theta=n\pi$),
\begin{equation}
    R_n-R \approx
    \frac{N'(\rp)}{4} \left[ \frac{r^2}{N^2(r)} \right]''_{\rp} \delta r_{\rm max}^2 e^{-\gamma n},
    \label{eq:Rn}
\end{equation}
which is identical to the spherically symmetric case up to the generalization of $\gamma$.

\begin{figure}
    \centering
    \includegraphics[width=\columnwidth]{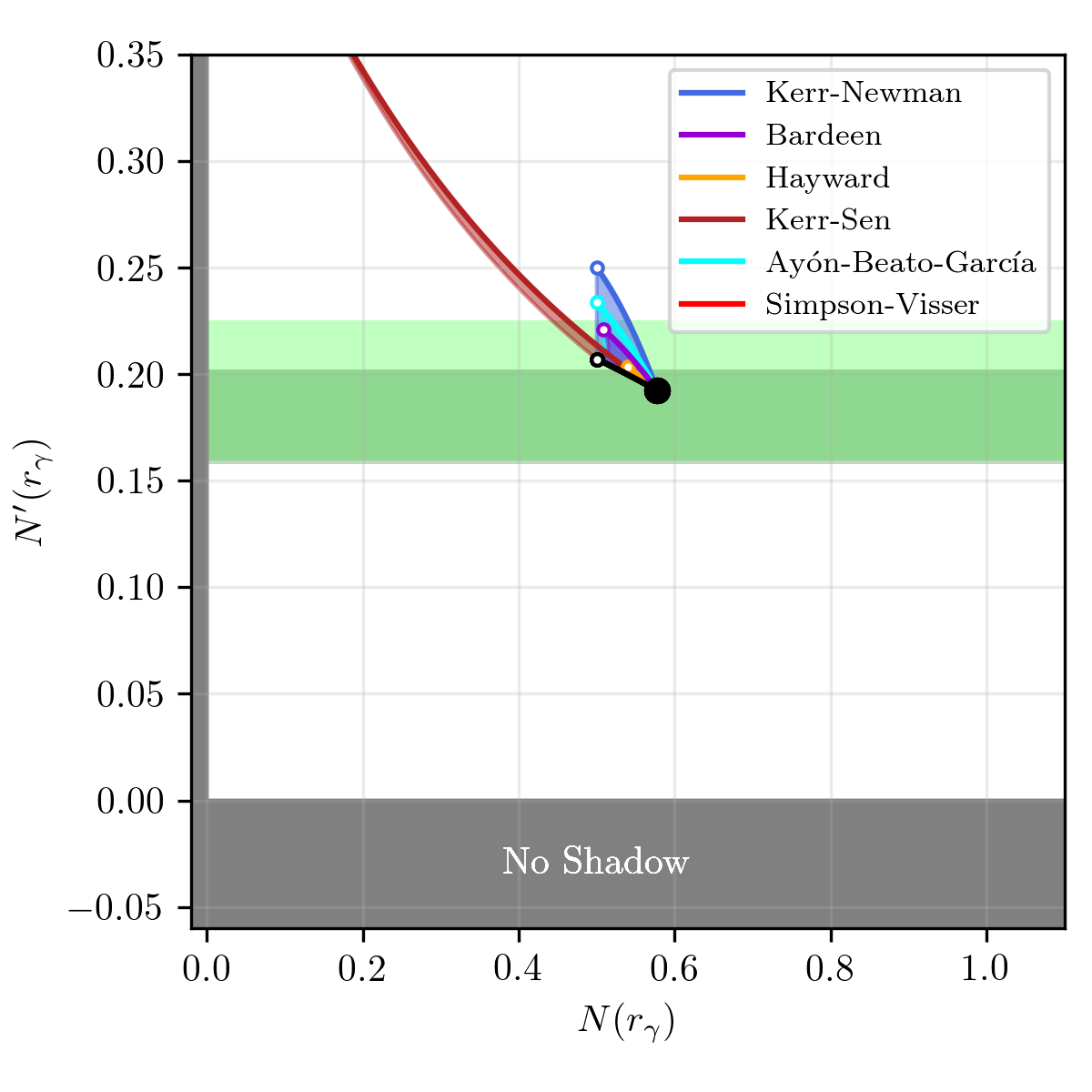}
    \caption{Comparison of 1$\sigma$ EHT constraints within the $N(\rp)$-$N'(\rp)$ plane arising from measurements of the size of the shadow with various spinning spacetimes (\citet{M87_PaperVI} in light green, $\theta_{n=1}$ in dark green, see \autoref{sec:measurements} for more information).  In each, the location of Schwarzschild is indicated by the black solid circular point, the path taken by Kerr is shown by the black line, terminating at $a=1$ (open black point).  For each alternative spinning spacetime, the range of values within the $N(\rp)$-$N'(\rp)$ plane spanned is shown by the shaded regions. The thick colored lines show the $a=0$ line spanned by the spacetime-specific charge, and terminate with an open point where the horizon ceases to exist.}
    \label{fig:NNp_comb}
\end{figure}

\begin{figure}
    \centering
    \includegraphics[width=\columnwidth]{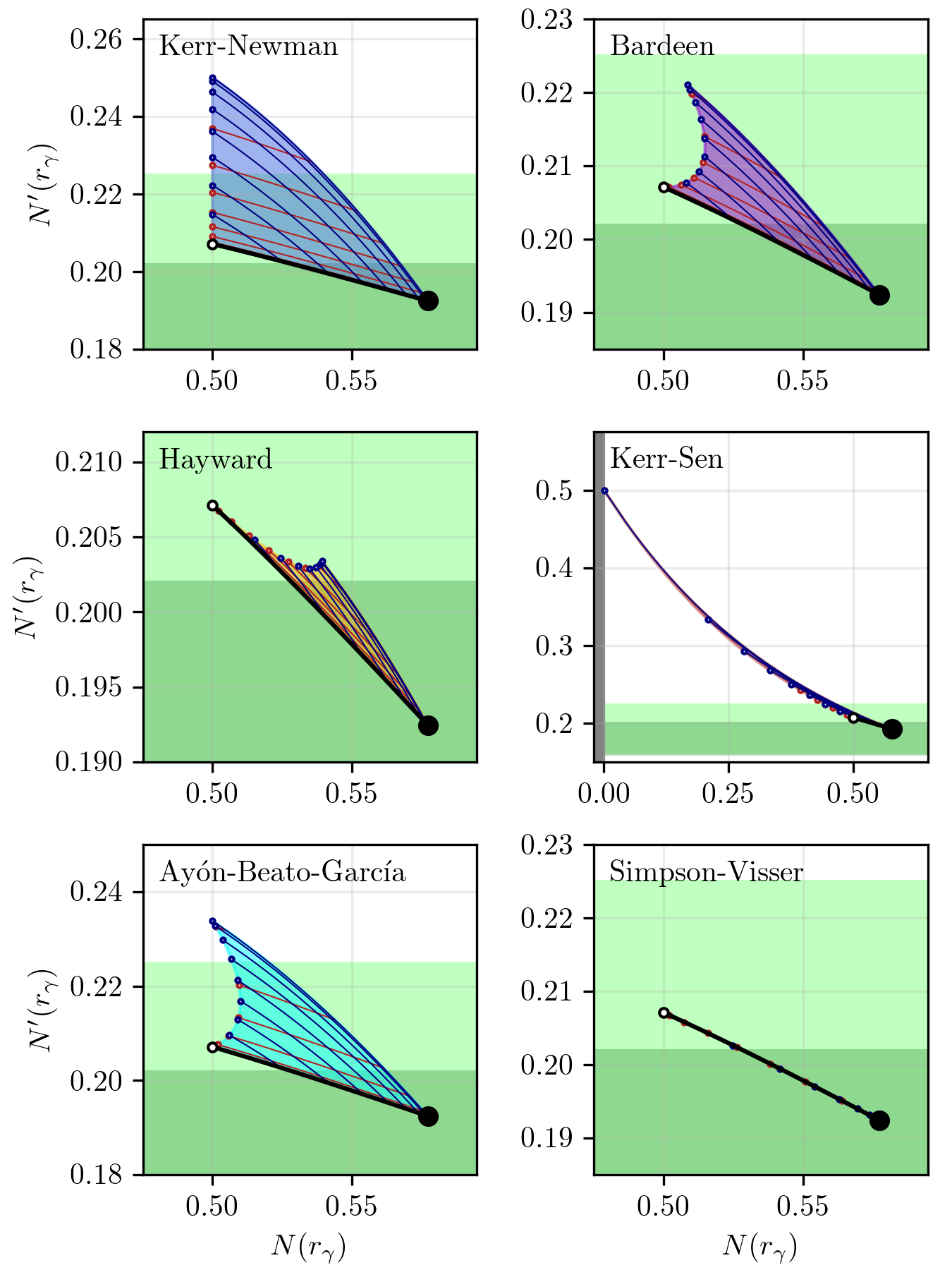}
    \caption{Same as \autoref{fig:NNp_comb} but zoomed in on the relevant region for each alternative spacetime. Red and blue lines shown constant charge and spin lines, respectively. Note that because the for the rotating Simpson-Visser spacetime (bottom right panel), $N(r)$ is the same as that in Kerr regardless of the additional charge, the shadow size is unchanged from the Kerr line.}
    \label{fig:NNp_multi}
\end{figure}

\begin{figure}
    \centering
    \includegraphics[width=\columnwidth]{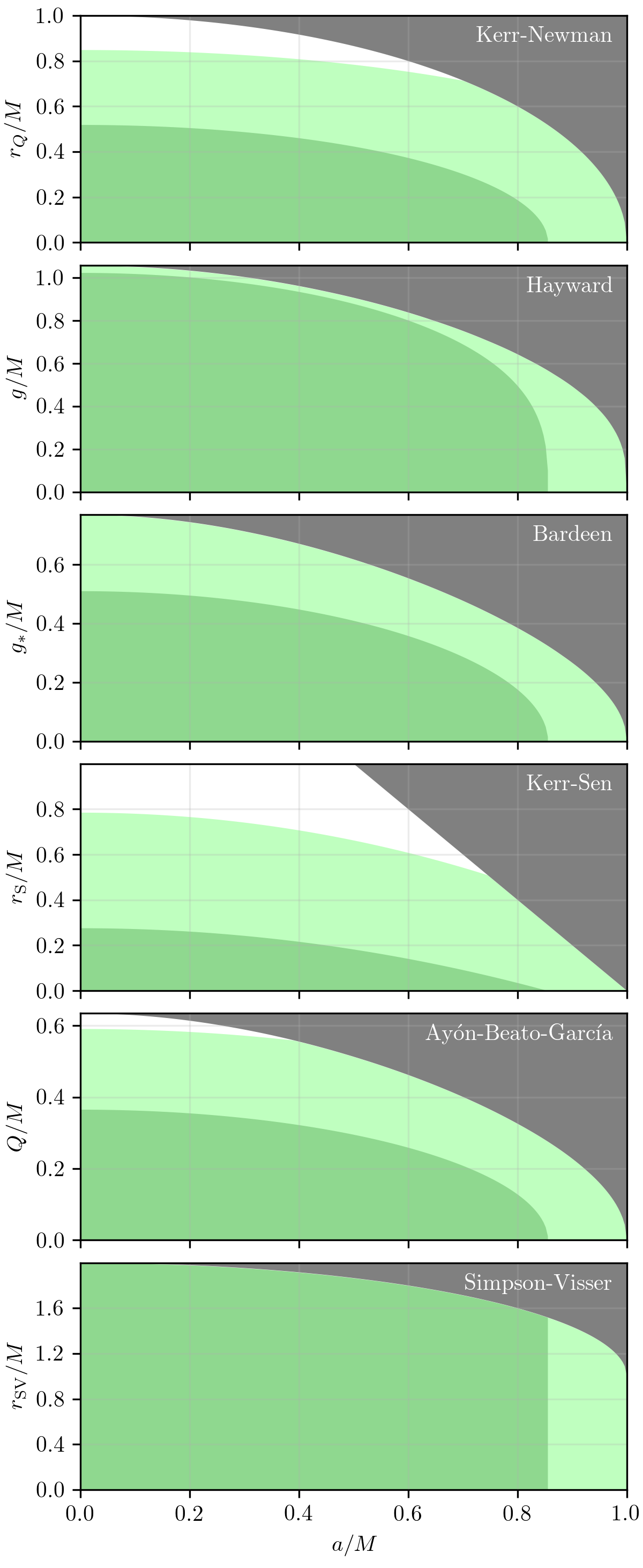}
    \caption{Constraints on charges for the various alternative spacetimes considered in \autoref{sec:obs}.  Dark and light green shaded regions indicate the 1$\sigma$ allowed regions from the EHT measurements shown in \autoref{fig:NNp_multi}.  The grey regions are excluded on theoretical grounds.}
    \label{fig:charges}
\end{figure}

\begin{deluxetable*}{lcccccccc}
\caption{Describing alternative spacetime metrics with \autoref{eq:metric} and the \citet{Johannsen2013} metric.}
\label{tab:altmetrics}
\tablehead{
\colhead{Metric/Family} & 
\colhead{$N^2(r)/F^2(r)$} &
\colhead{$B^2(r)/F^2(r)$} &
\colhead{$F(r)$} &
\colhead{$f(r)$} &
\colhead{$A^2_1(r)$} &
\colhead{$A^2_2(r)$} &
\colhead{$A_5(r)$}
}
\startdata
Kerr & 
$\Delta$ & 
$r^2$ &
$r/(r^2+a^2)$ & 
0 & 
1 & 
1 & 
1\\
Kerr-Newman & 
$\Delta_{\rm KN}$ &
$r^2$ &
$r/(r^2+a^2)$ & 
0 & 
$\Delta/\Delta_{\rm KN}$ &
$\Delta/\Delta_{\rm KN}$ &
$\Delta_{\rm KN}/\Delta$\\
Hayward & 
$\Delta_{\rm H}$ &
$r^2$ &
$r/(r^2+a^2)$ & 
0 & 
$\Delta/\Delta_{\rm H}$ &
$\Delta/\Delta_{\rm H}$ &
$\Delta_{\rm H}/\Delta$\\
Bardeen &
$\Delta_{\rm B}$ &
$r^2$ &
$r/(r^2+a^2)$ & 
0 &
$\Delta/\Delta_{\rm B}$ &
$\Delta/\Delta_{\rm B}$ &
$\Delta_{\rm B}/\Delta$\\
Kerr-Sen & 
$\Delta_{\rm S}$ &
$r^2$ &
$r/[r(r+r_{\rm S})+a^2]$ &
$r r_{\rm S}$ & 
$\displaystyle \frac{\Delta[r(r+r_{\rm S})+a^2]}{\Delta_{\rm S}(r^2+a^2)}$ &
$\Delta/\Delta_{\rm S}$ &
$\Delta_{\rm S}/\Delta$ \\
Simpson-Visser & 
$\Delta$ &
$r^2/\Delta_{\rm SV}$ &
$r/(r^2+a^2)$ &
$0$ & 
$1$ &
$1$ &
$\Delta_{\rm SV}$\\
Ay\'on-Beato-Garc\'ia & 
$\Delta_{AG}$ &
$r^2$ &
$r/(r^2+a^2)$ &
$0$ & 
$\Delta/\Delta_{AG}$ &
$\Delta/\Delta_{AG}$ &
$\Delta_{AG}/\Delta$\\
\\[-0.4cm]\hline\\[-0.4cm]
Johannsen &
$\displaystyle \frac{\Delta}{A_2^2(r)}$ &
$\displaystyle \frac{r^2}{A_2^2(r) A_5(r)}$ &
$\displaystyle \frac{r A_2(r)}{(r^2+a^2) A_1(r)}$ &
$f(r)$ &
$A_1^2(r)$ &
$A_2^2(r)$ &
$A_5(r)$ \\
Baines-Visser &
$\Delta_{BV} e^{-2\Phi}$ &
$r^2 e^{-2\Phi}$ &
$r/\Xi^2$ &
$\Xi-(r^2+a^2)$ &
$\displaystyle \frac{\Xi^2 \Delta e^{2\Phi}}{(r^2+a^2)^2 \Delta_{BV}}$ &
$\Delta e^{2\Phi} / \Delta_{\rm BV}$ &
$\Delta_{\rm BV}/\Delta$\\
Azreg-AÏnou &
$\Delta_{\rm A}$ &
$r^2 $ &
$r/(r^2+a^2)$ &
$f_{\rm A}(r)$ &
$\Delta/\Delta_{\rm A}$ &
$\Delta/\Delta_{\rm A}$ &
$\Delta_{\rm A}/\Delta$\\[0.25cm]
\enddata
\tablecomments{See \autoref{app:altmetrics} for details regarding coordinate and charge definitions.  The functions required to construct the \citet{Johannsen2013} metric as written in \autoref{eq:metric} are $N(r)$, $B(r)$, $F(r)$, and $f(r)$.  Alternatively, the functions necessary to construct the version in Equation~51 of \citet{Johannsen2013} are $A_1(r)$, $A_2(r)$, $A_5(r)$, and $f(r)$.}
\end{deluxetable*}

\section{Observational Implications}
\label{sec:obs}
Observational constraints on the shadow size and Lyapunov exponent translate via \autoref{eq:R} and \autoref{eq:gamma} into limits on $N(\rp)$, $N'(\rp)$, and $N''(\rp)/B^2(\rp)$.  We now explore these for recent EHT observations of \m87 and LIGO observations of the post-merger ringdown.  For explicit comparison, we include six alternative spacetimes -- Kerr-Newman \citep{Newman1965}, Rotating Hayward \citep{Hayward2006}, Rotating Bardeen \citep{Bardeen1973}, Kerr-Sen \citep{Sen}, Rotating Simpson-Visser \citep{SimpsonVisser2019,Shaikh2021}, and  Rotating Ay\'on-Beato-Garc\'ia \citep{AyonBeatoGarcia1998} -- each of which is described in more detail in \autoref{app:altmetrics}.  Each of these is characterized by a single parameter, or charge, in addition to spin that controls the perturbation from Kerr.  This selection is neither complete nor representative of the full range of potential set of axisymmetric, spinning perturbations.  They do, however, provide a familiar context in which to assess the description of the various measurements within the $N(\rp)$, $N'(\rp)$, and $N''(\rp)/B^2(\rp)$ presentation.

\subsection{Shadow Size Measurements}
\label{sec:measurements}
As noted previously, with an inclination of $17^\circ$, \m87 is viewed very nearly from the pole, justifying the approximation of a polar observer made here.  The otherwise unknown inclination of \sgra complicates the application of the polar shadow-size comparisons, and thus we focus our attention on \m87 alone.

The measurement of the bright emission radius is closely related to, but distinct from, the size of the shadow \citep[see, e.g.,][]{Narayan2019,Gralla2019}.  We will set aside the complications of relating EHT measurements of the ring-like emission in \m87 and the size of the shadow, and the associated astrophysical uncertainties; both of the measurements we consider here are calibrated using numerical accretion-flow simulations that are broadly successful in matching the multiwavelength properties of \m87 as well as its morphology at EHT wavelengths \citep{M87_PaperVI,M87_PaperV,MWL}.  We adopt the two implied \m87 shadow-size estimates as described in \citet{SphericalShadows23}. 

The first of these is based on the size of the emission region, calibrated to the mass via a set of numerical simulations.  Comparison of the reconstructed mass estimate with stellar dynamics observations results in an implied shadow size of,
\begin{equation}
    R/M
    = \sqrt{27} \left( 1.05_{-0.20}^{+0.15} \right).
\end{equation}

The second estimate is based on the extraction of the $n=1$ photon ring diameter by combined modeling and imaging.  The diameter of this strong-lensing feature is then calibrated to the shadow size via numerical simulations.  Again, comparing with stellar dynamics observations at larger distances, the shadow size estimate is,
\begin{equation}
    R/M = \sqrt{27} \left( 1.12_{-0.17}^{+0.10} \right).
\end{equation}
Because it is less dependent on the underlying astrophysical assumptions, the latter constraint on the size of the shadow has significantly smaller uncertainties, though it does rely critically on the interpretation of the two reconstructed ring and image components.  

Both shadow-size constraints limit $N'(\rp)$ alone and are shown in \autoref{fig:NNp_comb} and \autoref{fig:NNp_multi} as horizontal bands in the $N(\rp)$-$N'(\rp)$ plane.  In all plots the large black circle indicates the location of the Schwarzschild value, and is identical to that shown in \citet{SphericalShadows23}.  The black line shows the range of values spanned by Kerr, terminating at the open circle which indicates the maximal spin point ($a/M=1$).  The shadow size measurements are now competitive with the range of $N'(\rp)$ spanned by black hole spin, but as yet do not exclude any spins at 2$\sigma$.  

For each alternative spacetime example, lines of constant charge (red) and spin (blue) are shown in the zoomed regions on the right; filled points indicate maximal values at which an event horizon ceases to exist.  The regions spanned by the alternative spacetimes differ substantially in size and shape, indicating very different sensitivities to the EHT shadow size measurements.  Nevertheless, despite the significant heterogeneity in the specific form and motivation for the alternative spacetime, the observational implications are succinctly compared to each other and the EHT measurements.

The portion of the model-regions that lie inside of the EHT measurement band are preferred, and conversely those outside of the green bands are unpreferred.  On a spacetime-by-spacetime basis, these may be translated into limits on the additional charge, shown in \autoref{fig:charges}.  Currently, the only alternatives for which the EHT shadow size results in a significant (2$\sigma$) constraint on the additional charge are the Kerr-Newman, Kerr-Sen, and Ay\'on-Beato-Garc\'ia models, consistent with the conclusions of \citet{Prashant2021}.  Note that because for the Simpson-Visser spacetime, $N(r)$ takes the same form as for Kerr, the measurement of a shadow size presents no constraint on the additional charge.

\subsection{Lyapunov Exponent Measurements}

\begin{figure}
    \centering
    \includegraphics[width=\columnwidth]{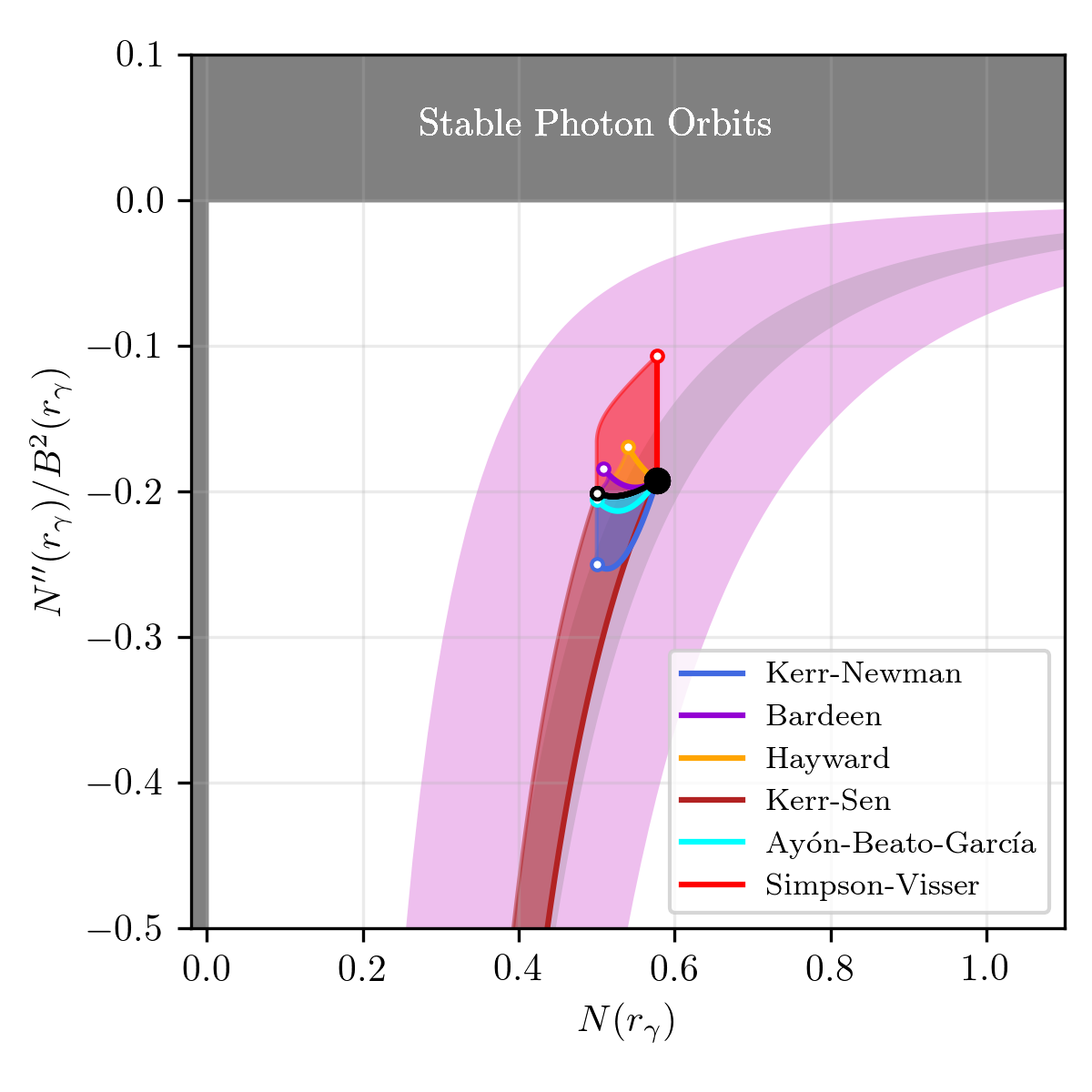}
    \caption{
    Implications of measurements of the Lyapunov exponent in the $N(\rp)$-$N''(\rp)/B^2(\rp)$ plane..  The light grey band indicates the constraning power of a 10\% measurement of $\gamma$.  The implied 1$\sigma$ region associated with the 220 QNM frequency and damping rate measurements reported in \citet{LIGOQNM} is shown by the magenta band.  In each panel, the location of Schwarzschild is indicated by the black solid circular point, the path taken by Kerr is shown by the black line, terminating at $a=1$ (open black point).  For each alternative spinning spacetime listed in \autoref{sec:obs}, the range of values within the $N(\rp)$-$N''(\rp)/B^2(\rp)$ plane spanned is shown by the shaded regions.  The thick colored lines show the $a=0$ line spanned by the spacetime-specific charge, and terminate with an open point where the horizon ceases to exist.}
    \label{fig:NNpp_comb}
\end{figure}

\begin{figure}
    \centering
    \includegraphics[width=\columnwidth]{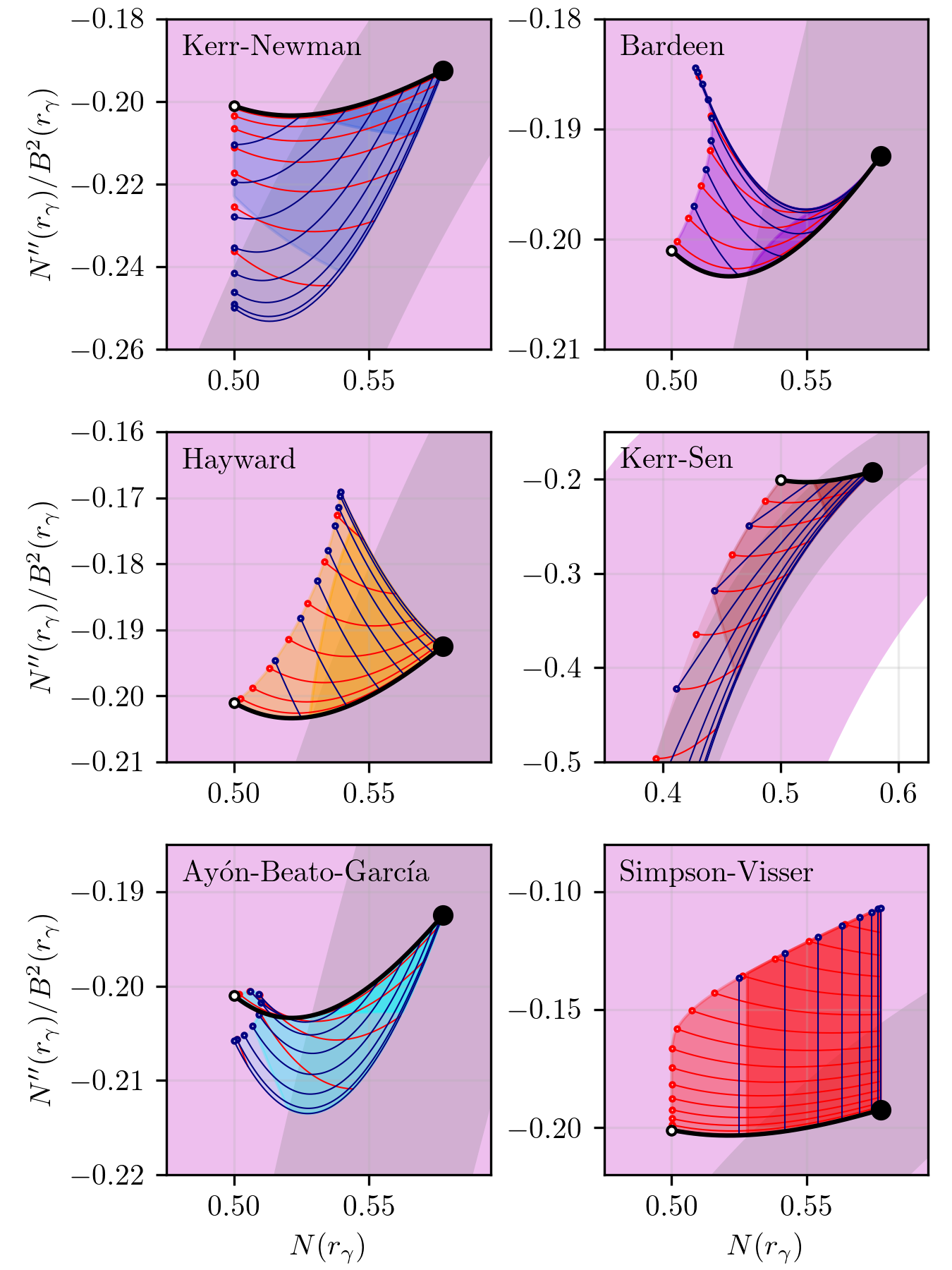}
    \caption{Same as \autoref{fig:NNpp_comb} but zoomed in on the relevant region for each alternative spacetime. Red and blue lines shown constant charge and spin lines, respectively. The darkness of the shading indicates the 1$\sigma$ allowed regions based on the EHT shadow size estimates shown in \autoref{fig:NNp_multi} and \autoref{fig:charges}.}
    \label{fig:NNpp_multi}
\end{figure}

In \citet{SphericalShadows23}, it was shown that the measurement of multiple photon rings provided a novel way to measure $\gamma$.  This remains true in the case of spinning spacetimes, as implied by studies of the value of even low-order photon ring measurements \citep{Spin}.  We make this concrete via repeated application of \autoref{eq:Rn}.  Even without an explicit value for $\delta r_{\rm max}$, the relative displacements of the of subsequent photon rings from the shadow may be related to the Lyapunov exponent alone:
\begin{equation}
    \frac{R_n-R}{R_{n+1}-R}= e^{\gamma}.
\end{equation}
Thus, measuring the shadow size and two photon ring radii in M87* may be transformed into a measurement of $\gamma$.  The need to measure the shadow size may be eliminated altogether if three photon ring radii can be measured,
\begin{equation}
    \gamma = \ln \left(\frac{R_{n+1}-R_n}{R_{n+2}-R_{n+1}}\right).
\end{equation}
\autoref{fig:NNpp_comb} and \autoref{fig:NNpp_multi} illustrate the implications of a measurement of $\gamma$ with 10\% accuracy (shaded light-grey region), in comparison to the physically allowed range of the various explicit alternative metrics.  In most cases, this imposes a nearly orthogonal constraint to the measurement of the shadow size.
In particular, the Rotating Simpson-Visser spacetime is no longer degenerate with Kerr due to the differing $B(r)$ (see \autoref{tab:altmetrics} and \autoref{app:altmetrics}).

 \begin{figure}
     \centering
     \includegraphics[width=\columnwidth]{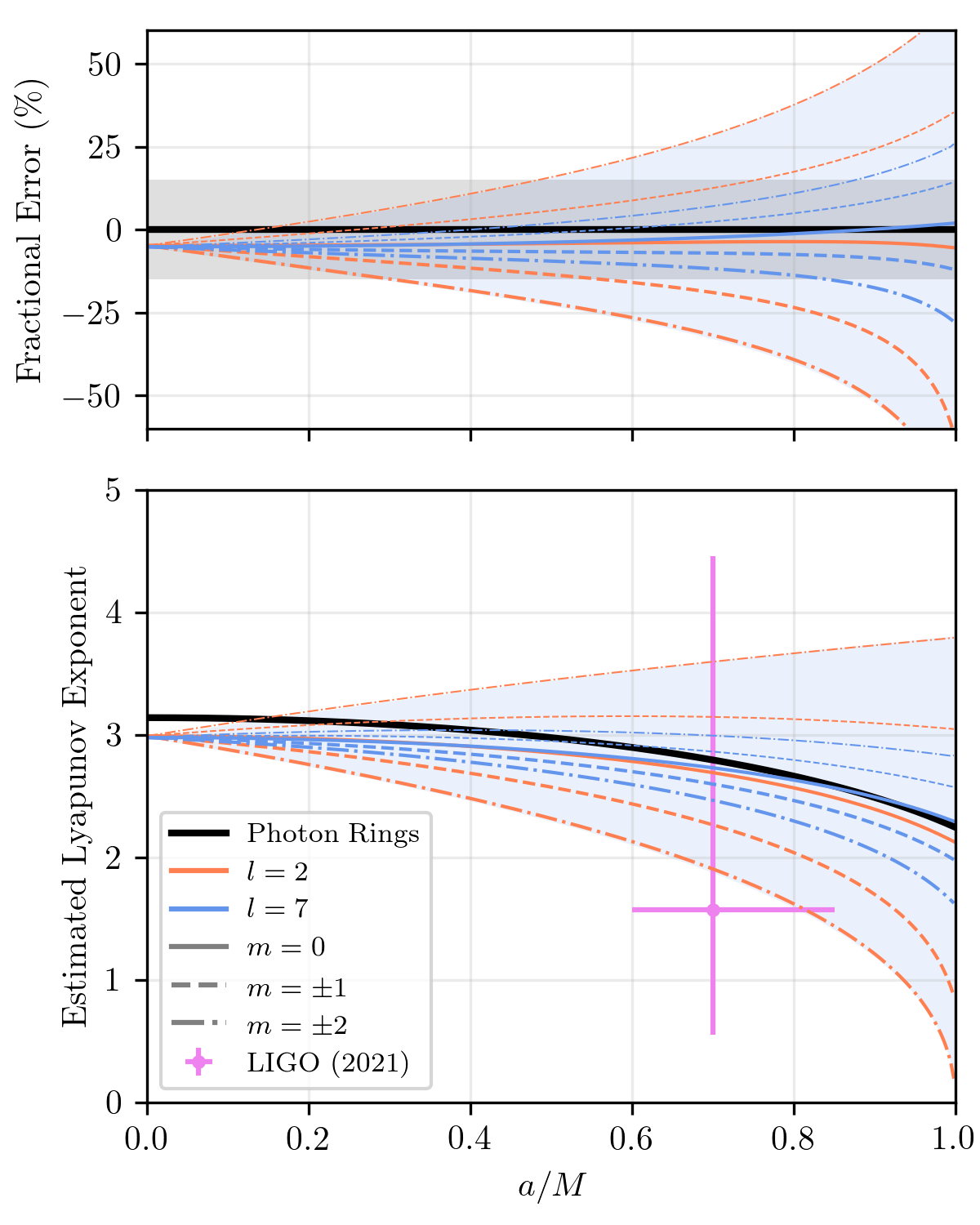}
     \caption{Bottom: polar-orbit Lyapunov exponent from \autoref{eq:gamma} (black) in comparison to the estimates from the fundamental ($n=0$) quasinormal modes, $2\pi l \Im(\omega_{lmn})/\Re(\omega_{lmn})$ for various $l$ and $m$ as tabulated by \citet{Berti2009}.  The range of values for $l=7$ encompassed by $|m|\le l$ is shaded in blue, with those for $m=-1,-2$ and $m=0,1,2$ shown by the thin and thick lines, respectively, for $l=2,7$.  For reference, the implied estimates from \citet{LIGOQNM} are shown in magenta (error bars indicate 90\% $4\pi\omega_I/\omega_R$ and final spin ranges).  Top: the fractional difference between the QNM estimates and the polar-orbit Lyapunov exponent.  The light grey band shows the 15\% region, similar to those expected for high S/N LIGO events \citep{IsiFarr2021}.}
     \label{fig:gamma_qnm}
 \end{figure}

In principle, black hole mergers provide an alternative way to directly measure $\gamma$.  The late-time gravitational wave signal is dominated by the oscillation and dissipation of the quasinormal modes (QNMs) of the product black hole.  For high angular momentum modes, the frequency and damping timescales are related to the Lyapunov exponent near the photon orbit \citep[see, e.g.,][]{Huan2021}, i.e.,
\begin{equation}
    \lim_{l\rightarrow\infty} \frac{2\pi l \omega_{I,lm0}}{\omega_{R,lm0}} = \gamma,
\end{equation}
where $\omega_{lmn} = \omega_{R,lmn} + i \omega_{I,lmn}$ is the complex frequency of the QNM with angular quantum numbers $l$ and $m$, measured relative to the polar axis, and radial quantum number $n$ (i.e., $n=0$ is the fundamental).  As these are dominated by the dynamics of massless fields near the horizon, this behavior is unsurprising.

In practice, $2\pi l \omega_{I,lm0}/\omega_{R,lm0}$ approaches $\gamma$ very rapidly, and for Kerr is within 5.5\% for the $lmn=200$ mode (polar quadrupolar mode) at all values of $a$, shown in \autoref{fig:gamma_qnm}.  The more commonly discussed $lmn=220$ mode (azimuthal quadrupolar mode) exhibits larger departures from the $\gamma$ measured by a polar observer, and is almost certainly a consequence of the distinction between the polar and equatorial photon trajectories and their associated Lyapunov exponents.  (For Kerr, the 220 and 770 modes have nearly identical $2\pi l\omega_{I,lm0}/\omega_{R,lm0}$, strongly suggesting that value of $\gamma$ associated with azimuthally propagating modes is responsible for the departure from \autoref{eq:gamma}.)  We leave the investigation of the Lyapunov exponents for arbitrary viewing inclinations for future work.

The focus of the gravitational wave community on the 220 mode is driven by two considerations within the Kerr context.  First, the damping times of higher-$l$ and higher-$n$ modes are substantially shorter, and thus are unlikely to be among the first QNMs constrained.  The difference in damping times among different azimuthal quantum numbers, $m$, are much smaller; $\omega_{I,220}/\omega_{I,200}>0.9$ for $a\le0.8$, falling below $0.5$ only for $a>0.98$.  Thus, even at similar S/N, the polar quadrupolar mode should be detectable.

Second, the QNMs that are relevant for LIGO and other gravitational wave detectors are presumably excited by the residuals of the binary orbital motions prior to merger.  When the spin of the final merger product is parallel to the binary orbital angular momentum, these motions preferentially excite the azimuthal quadrupolar modes, and by symmetry cannot excite the polar quadrupolar mode.  However, when the spin of the final merger product is misaligned with the orbit \citep[via, e.g., the mechanisms described in][]{SteinleKesden2021,SuLai2021}, it is possible to excite the full compliment of $l=2$ modes, including the polar quadrupolar mode, albeit subdominently.

Nevertheless, for the final spin estimates of LIGO events for which QNM measurements have been made, the range of reconstructed $\gamma$ are consistent with all quadrupolar $m$, shown explicitly in \autoref{fig:gamma_qnm} for the joint estimates on the deviation from the Kerr 220 mode \citep{LIGOQNM}.  Future events with high S/N should permit estimates with precisions as high as 15\% \citep{IsiFarr2021}, and thus will be able to distinguish the polar (200) and azimuthal (220) quadrupolar modes.  \autoref{fig:NNpp_comb} and \autoref{fig:NNpp_multi} show the 1$\sigma$ region from the \citet{LIGOQNM} QNM constraints.

\section{Conclusions}
\label{sec:conclusions}

The nonperturbative and nonparametric formalism for characterizing the gravitational implications of near-horizon observations, including the detection of a black hole shadow, proposed in \citet{SphericalShadows23} is applicable for a large class of axisymmetric, spinning spacetimes.  As in the spherically symmetric case, this formalism renders explicit what gravitational features are, and are not, constrained by the various classes of measurement.  We find that for polar observers, an appropriate initial approximation for \m87, the conclusions previously obtained for general spherically symmetric spacetimes may be naturally extended to the broad class of rotating spacetimes described by \citet{Johannsen2013}.

For our purposes, we have reformulated the large family of stationary, axisymmetric and integrable metrics proposed by \citet{Johannsen2013}, in terms of four unknown functions of radius that explicitly identify and separate those responsible for setting the location of the event horizon, $N(r)$, the boundary of the ergosphere, $F(r)$.  The ability to independently set these locations permits the construction of a wide variety of alternative spacetimes of practical interest that isolate the impact of these key dynamical locations within the spacetime.

We explicitly demonstrate that the \citet{Johannsen2013} metric is equivalent to or encompasses a number of other metric families, including those proposed by \citep{Visser2023,AzregAinou2014}, and the empirically relevant subset from \citet{CarsonYagi2020}.  While not a proof of universality, this commonality suggests that our form of the \citet{Johannsen2013} metric comprises a broad class of theoretically interesting alternative spacetimes.  This interpretation is supported by the fact that many alternative black hole metrics may be immediately expressed in this form, summarized in \autoref{tab:altmetrics}.  A notable exception is the spacetime proposed in \citet{Rasheed1995}, which while stationary and axisymmetric is not integrable.

The size of the shadow seen by polar observers is directly related to $N'(\rp)$, as found in the \citet{SphericalShadows23}: measuring the shadow size is roughly synonymous with measuring the radial derivative of $g_{tt}$ along the polar axis in Boyer-Lindquist coordinates.  In this sense, precise shadow size measurements generate precise metric constraints.  However, there is no constraint on $N(\rp)$ from the shadow size alone, and thus shadow size measurements by themselves do not provide any limit on the magnitude of a putative deviation in $g_{tt}$.  In this sense, precise shadow measurements are uninformative.  

Shadow measurements do not offer any insights into the various other arbitrary functions of the general metric, $F(r)$, $B(r)$, or $f(r)$, and therefore, do not provide information about the location or properties of the ergosphere, both of which are set by $F(r)$.  As a consequence, extracting lensing features from horizon-resolving observations like those of \m87 by EHT or subsequent missions, may not provide purely gravitational tests of frame dragging by astrophysical black holes in the near future.  Of course, we cannot preclude the possibility of astrophysical modeling providing such insights through the dynamical impact of the ergosphere on the matter and electromagnetic fields in the black hole's vicinity.

Observations of high-order photon rings would provide direct information about the Lyapunov exponent of circular orbits near the polar photon sphere, and therefore on $N(\rp)$, $N'(\rp)$, and $N''(\rp)/B^2(\rp)$. 
The expression that we find for $\gamma$ matches that in \citet{SphericalShadows23} for $a=0$ exactly, and differs by a spin-dependent factor of order unity for all $a/M$ except those within $10^{-7}$ of near $1/N'(\rp)$; for maximally spinning Kerr this factor is 1.059.  As with shadow size measurements, observations of high-order photon rings by a polar observer do not constrain $F(r)$, and therefore do not probe the ergosphere.  

The nonparametric and nonperturbative framework for characterizing near-horizon gravity tests proposed in \citet{SphericalShadows23} -- expressing these as constraints in the $N(\rp)$-$N'(\rp)$-$N''(\rp)/B^2(\rp)$ volume -- may be used in the context of a very broad class of spinning spacetimes.  As for the spherically symmetric case, expressing constraints in this way directly avoids concerns regarding specific choices of parameterizations (with their attendant prior assumptions), the nonlinearity of gravity near the horizon, or the need for explicit alternative gravity theories.

We apply this framework to EHT observations of \m87, which due to the low inclination are well approximated by those by a polar observer.  As with the spherically symmetric case, the detection of a shadow carries both qualitative and quantitative implications.  The mere detection of a shadow implies that at the outermost polar photon orbit, $N'(\rp)>0$.  Reported EHT measurements of the radius of the shadow result in direct constraints on $M/N'(\rp)$.  For specific alternative spacetimes, these measurements translate into constraints on the additional charges.  Of the alternatives we consider, 2$\sigma$ constraints are found on the charges of the Kerr-Newman ($r_Q/M<0.91$), Kerr-Sen ($r_{\rm S}/M<0.92$), and Ay\'on-Beato-Garc\'ia ($Q/M<0.63)$ spacetimes.  These limits are spin dependent, generally becoming stronger at higher spins.

Similarly, the observation of a photon ring results in important qualitative and quantitative conclusions.  The instability of photon trajectories on the polar photon sphere requires $N''(\rp)<0$, and thus the detection of even a single order photon ring implies this is true.  In the future, measurements of the radii of multiple photon rings may present an opportunity to directly extract the Lyapunov exponent of trajectories near the polar photon sphere.  Existing gravitational wave observations of the late-time ringdown of black hole merger events by LIGO provide a second way to measure the Lyapunov exponent.  While at present, the precision with which the QNM frequency and damping rate is insufficient to significantly constrain any of the alternative black hole spacetimes we consider, the precision permitted by future gravitational wave detectors will do so.  At such time, it will become important to distinguish between modes with diffreent azimuthal quantum numbers, with the $m=0$ mode frequencies being most relevant for direct comparison with EHT observations of \m87.

A significant limitation of our work is the restriction to a polar observer.  While an appropriate initial approximation for \m87, it is not obviously applicable to \sgra, for which the inclination is currently uncertain.  Extension to non-polar observers may create opportunities to probe the ergosphere, a possibility supported by the spin-dependence of the Lyapunov exponent estimate from QNMs with $m\ne0$.  We leave this extension for future work.  

Additionally, while extremely broad, the class of axsymmetric spacetimes that we considered is not general.  It excludes nonintegrable spacetimes, of which examples exist in the literature.  It further excludes nonstationary spacetimes, though this restriction is unlikely to be a serious limitation for imaging experiments like those performed by EHT and future instruments.

We have not made any effort to critically assess the connection of the purely gravitational lensing features addressed here to practical, astrophysically-dependent observations.  That is, we address neither the difficulties inherent in measuring the radius of the shadow nor the astrophysical uncertainties associated with measuring the radii of high-order photon rings \citep[see, e.g.][]{Spin}.  Both are relevant insofar as they limit the precision with which inferences about strong gravity can be made, and will be addressed elsewhere.

Finally, we have focused our attention solely on lensing signatures (of which QNMs are also an example despite the dynamical nature of gravitational waves).  However, astrophysical modeling provides the opportunity for high-precision gravitational insights obtained in part via the impact of spacetime structure on the dynamics of particles and fields near the black hole \citep[e.g.][]{Broderick2014,Johannsen2016,Tiede2020}.

\begin{acknowledgments}
\section{acknowledgments}
The Perimeter Institute for Theoretical Physics partially supported this work. Funding for research at the institute is provided by the Department of Innovation, Science and Economic Development Canada, and the Ministry of Economic Development, Job Creation and Trade of Ontario, both of which are branches of the Government of Canada. A.E.B. expresses gratitude to the Delaney Family for their generous financial backing through the Delaney Family John A. Wheeler Chair at Perimeter Institute. Additionally, A.E.B. receives further financial support for this research through a Discovery Grant from the Natural Sciences and Engineering Research Council of Canada.
\end{acknowledgments}

\vspace{5mm}

\appendix

\section{ADM Mass and Spin}
\label{app:asympADM}
The parameters $M$ and $a$ correspond to the ADM mass and spin, respectively, when $N(r)$, $B(r)$, $F(r)$, and $f(r)$ approach their Kerr values at large $r$.  That is, should it be desirable to interpret $M$ and $a$ as the ADM mass and spin, this imposes the following asymptotic conditions on the otherwise unrestricted functions,
\begin{equation}
\begin{gathered}
    \lim_{r\rightarrow\infty} f(r) = 0\\
    \lim_{r\rightarrow\infty} r[N(r)-(1-M/r)] = 0\\
    \lim_{r\rightarrow\infty} r[B(r)-1] = 0\\
    \lim_{r\rightarrow\infty} r[F(r)-1] = 0.
\end{gathered}
\end{equation}

\newpage
\section{Metric Components}
\label{app:metric_comp}
The components of the metric in \autoref{eq:metric} may be written as
\begin{equation}
\begin{aligned}
    g_{tt} &= 
    -\tilde{\Sigma} \frac{N^2-a^2 F^2 \sin^2\theta}{[r-a^2F\sin^2\theta]^2} \\
    g_{rr} &= \frac{\tilde{\Sigma}}{r^2} \frac{B^2}{N^2} \\
    g_{\theta\theta} &= \tilde{\Sigma} \\
    g_{\phi\phi} &= \tilde{\Sigma}\sin^2\theta \frac{[r^2-a^2N^2\sin^2\theta]}{[r-a^2F\sin^2\theta]^2}\\
    g_{t\phi} &= \tilde{\Sigma} a\sin^2\theta 
    \frac{[N^2-rF]}{[r-a^2F\sin^2\theta]^2}.
\end{aligned}
\label{eq:metric_components}
\end{equation}
From these we have
\begin{equation}
    g_{t\phi}^2-g_{tt}g_{\phi\phi}
    =
    \frac{\tilde{\Sigma}^2 N^2 \sin^2\theta}{[r-a^2F\sin^2\theta]^2},
\end{equation}
and therefore, the metric determinant is
\begin{equation}
    \sqrt{-g}
    =
    \frac{\tilde{\Sigma}^2 B}{r-a^2F\sin^2\theta}.
\end{equation}

With the identifications,
\begin{equation}
\begin{aligned}
    A_1^2(r) &= \frac{r^2\Delta}{N^2(r) (r^2+a^2)^2}\\
    A_2^2(r) &= \frac{F^2(r) \Delta}{N^2(r)}\\
    A_5(r) &= \frac{r^2 N^2(r)}{\Delta B^2(r)},
\end{aligned}
\end{equation}
and $f(r)$ is unchanged, these match Equation~51 of \citet{Johannsen2013} exactly.  For reasons specified in \autoref{sec:metric}, we adopt the parameterization in terms of the unknown functions $N(r)$, $B(r)$, $F(r)$, and $f(r)$.

\section{The Polar Photon Sphere}
\label{app:rp}
We begin with \autoref{eq:j13_polar_eom} for $\dot{r}^2$,
\begin{equation}
    \dot{r}^2 = 
    \frac{r^4}{\tilde{\Sigma}^2}
    \frac{e^2}{B^2(r)} \left[
    1 - \frac{q^2+a^2}{r^2} N^2(r)
    \right].
    \tag{\ref{eq:j13_polar_eom}}
\end{equation}
To obtain an equation for $\ddot{r}$, we differentiate with respect to the affine parameterization of nearby geodesics ($\lambda$) and take the limit as $\dot{r}\rightarrow0$.  That is,
\begin{equation}
\begin{aligned}
    \ddot{r}
    &= 
    \frac{1}{2\dot{r}}\frac{d\dot{r}^2}{d\lambda}\\
    &=
    \frac{1}{2\dot{r}}
    \left(
        \frac{\partial\dot{r}^2}{\partial r} \dot{r}
        +
        \frac{\partial\dot{r}^2}{\partial \theta} \dot{\theta}
    \right)\\
    &=
    \frac{1}{2}
    \left[
    \frac{r^4}{\tilde{\Sigma}^2}
    \frac{e^2}{B^2(r)} 
    \right]'
    \left[
    1 - \frac{q^2+a^2}{r^2} N^2(r)
    \right]\\
    &\qquad
    -\frac{1}{2} \frac{r^4}{\tilde{\Sigma}^2}
    \frac{e^2}{B^2(r)}
    \left[
    \frac{q^2+a^2}{r^2} N^2(r)
    \right]'\\
    &\qquad
     +\frac{a^2\sin(2\theta)}{\tilde{\Sigma}} \dot{r} \dot{\theta}.
\end{aligned}
\end{equation}
Taking the limit as $r \rightarrow \rp$, and therefore $\dot{r}\rightarrow 0$ and $r^2\rightarrow (q^2+a^2) N^2(r)$, the first and third terms vanish identically.  This leaves
\begin{equation}
    \ddot{r} = 
    - \frac{r^4}{\tilde{\Sigma}^2}
    \frac{e^2}{B^2(r)}
    \left[
    \frac{q^2+a^2}{r^2} N^2(r)
    \right]'
\end{equation}
which we assume vanishes only when the second term does, i.e., $[(q^2+a^2)N^2(r)/r^2]'=0$.  Thus, we have two simultaneous equations for $r$ and $q$, which combine to give the desired algebraic expression, $\rp N'(\rp) = N(\rp)$.

For the Kerr spacetime, $N^2(r) = \Delta r^2/(r^2+a^2)^2$ (see \autoref{app:altmetrics}), and the equation for the polar photon sphere reduces a polynomial,
\begin{equation}
    \rp^3-3M\rp^2+a^2\rp + a^2 M = 0.
    \label{eq:Kerr_rppoly}
\end{equation}
In the limit that $a=0$, the sole non-trivial solution is $\rp=3M$.  For arbitrary spins, subject to $|a|<M$, this polynomial admits three solutions, only one of which is generally real valued:
\begin{equation}
\begin{split}
     \rp & = M + 2M \sqrt{1-\frac{a^2}{3M^2}}\\ & \times
    \cos\left[
    \frac{1}{3} \cos^{-1}\left(
    \frac{1-a^2/M^2}{(1-a^2/3M)^{3/2}}
    \right)
    \right].
    \label{eq:Kerr_rp}
\end{split}  
\end{equation}
Note that these do not match the prograde/retrograde equatorial photon sphere radii, i.e., those obtained by inspecting only null geodesics confined to the equatorial plane (see, e.g., BARDEEN 1972).  Indeed, as must be the case, the polar photon sphere radius is even in $a$, while the radius of the equatorial photon sphere depends on the relative direction of the spin and orbital angular momentum.

\section{Describing Alternative Spacetimes}
\label{app:altmetrics}
The metric in \autoref{eq:metric} is sufficiently general to describe a wide range of spinning black hole spacetimes.  Here we explicitly demonstrate that it can express a number of specific alternative spacetimes by constructing the associated free functions.

\subsection{Kerr}
We begin with the Kerr metric, written in Boyer-Lindquist coordinates,
\begin{equation}
    \begin{split}
         ds^2 &= - \frac{\Delta}{\Sigma} (dt - a \sin^2\theta d\phi)^2\\ &
    + \frac{\sin^2\theta}{\Sigma} \left[ a dt - (r^2+a^2) d\phi \right]^2\\ &
    + \frac{\Sigma}{\Delta} dr^2 + \Sigma d\theta^2,
    \label{eq:metric_kerr}
    \end{split}
\end{equation}

where $\Sigma=r^2+a^2\cos^2\theta$.  This form may be immediately represent in the form of \autoref{eq:metric} with appropriate choices for $N(r)$, $F(r)$, $B(r)$, and $f(r)$.  Before explicitly stating these (see also \autoref{tab:altmetrics}), we begin with define a function that will appear repeatedly, $u(r) \equiv r/(r^2+a^2)$, and note that when $F(r)=u(r)$,
\begin{equation}
    r-F a ^2 \sin^2\theta = \Sigma u(r),
\end{equation}
i.e., with an appropriate definition of $F(r)$ some of the denominators in \autoref{eq:metric} (or \autoref{eq:metric_components}) simplify in such a way that the dependence on the polar angle may be subsumed into factors of $\Sigma$.  Therefore, inserting this expression for $F(r)$ into \autoref{eq:metric} produces
\begin{equation}
    \begin{split}
         ds^2 &
    =
    -\frac{\tilde{\Sigma} N^2}{\Sigma^2 u^2(r)} (dt-a\sin^2\theta d\phi)^2\\ &
    +
    \frac{ \tilde{\Sigma}\sin^2\theta}{\Sigma^2} \left[ a dt - (r^2+a^2) d\phi \right]^2\\ &
    +
    \frac{\tilde{\Sigma}}{r^2} \frac{B^2}{N^2} dr^2
    +
    \tilde{\Sigma} d\theta^2.
    \label{eq:metric_Fsimplified}
    \end{split}
\end{equation}

The remaining functions in \autoref{eq:metric} are set by matching the coefficients with those in \autoref{eq:metric_kerr}.  We begin with setting the coefficient of $d\theta^2$, which requires $f(r)=0$ and thus $\tilde{\Sigma}=\Sigma$.  This choice also results the matching of the coefficient in the second term.  The first term may be used to set $N(r)$,
\begin{equation}
    N^2(r) = \Delta u^2(r),
\end{equation}
and the third sets $B(r)$,
\begin{equation}
    B^2(r) = \frac{r^2}{\Delta} N^2(r) = r^2 u^2(r).
\end{equation}
These are summarized in \autoref{tab:altmetrics}.

\subsection{Kerr-Newman}
Charged spinning black holes are described in general relativity by the Kerr-Newman metric,
\begin{equation}
    \begin{split}
        ds^2 & = - \frac{\Delta_{\rm KN}}{\Sigma} (dt - a \sin^2\theta d\phi)^2\\ &
    + \frac{\sin^2\theta}{\Sigma} \left[ a dt - (r^2+a^2)   d\phi \right]^2\\  &
    +  \frac{\Sigma}{\Delta_{\rm KN}} dr^2 + \Sigma d\theta^2,
    \label{eq:metric_kerrnewman}
    \end{split}
\end{equation}
which differs from Kerr by the introduction of $\Delta_{\rm KN}$,
\begin{equation}
    \Delta_{\rm KN}
    \equiv
    r^2 - 2Mr + a^2 + r_Q^2,
\end{equation}
where $r_Q^2\equiv GQ^2/4\pi\epsilon_0 c^4$ is the black hole charge in gravitational units \citep{Newman1965}.  Again, we set $F(r)=u(r)$, which puts the metric in the form of \autoref{eq:metric_Fsimplified}.  Then, we adopt
\begin{equation}
\begin{aligned}
    N^2(r) &= \Delta_{\rm KN} u^2(r) \\
    B^2(r) &= r^2 u^2(r) \\
    f(r) &= 0,
\end{aligned}
\end{equation}
for which \autoref{eq:metric_Fsimplified} becomes identical to \autoref{eq:metric_kerrnewman}.  These are summarized in \autoref{tab:altmetrics}.

As with Kerr, the equation for the polar photon orbit sphere radius reduces to a cubic polynomial,
\begin{equation}
    \rp^3 - 3M\rp^2 + (2r_Q^2+a^2) r + a^2 M = 0,
\end{equation}
which is very similar to \autoref{eq:Kerr_rppoly}.  
The associated $\rp$ is
\begin{equation}
    \begin{split}
        \rp & = M + 2M \sqrt{1-\frac{(a^2+2r_Q^2)}{3M^2}}\\ &
    \times \cos\left\{
    \frac{1}{3} \cos^{-1}\left[
    \frac{1-(a^2+r_Q^2)/M^2}{[1-(a^2+2r_Q^2)/3M]^{3/2}}
    \right]
    \right\},
    \label{eq:KerrNewan_rp}
    \end{split}
\end{equation}

which is very similar to \autoref{eq:Kerr_rp}.  The shadow size may be immediately computed via \autoref{eq:R}.

\subsection{Rotating Hayward}
The Hayward metric is originally spherically symmetric, and thus a substitute for Schwarzschild \citep{Hayward2006}.  A rotating analog has been constructed via a procedure analogous to the Newman-Janis algorithm, resulting in a Kerr-like metric with mass replaced by a function of radius (Abdujabbarov et al. 2016, though see Bambi and Modesto, 2013).  Expressed in Boyer-Lindquist coordinates, this takes the form
\begin{equation}
    \begin{split}
        ds^2 & = - \frac{\Delta_{\rm H}}{\Sigma} (dt - a \sin^2\theta d\phi)^2\\ &
    + \frac{\sin^2\theta}{\Sigma} \left[ a dt - (r^2+a^2) d\phi \right]^2\\ &
    + \frac{\Sigma}{\Delta_{\rm H}} dr^2 + \Sigma d\theta^2,
    \label{eq:metric_hayward}
    \end{split}
\end{equation}

where
\begin{equation}
    \Delta_{\rm H} = r^2 + a^2 - 2\frac{M r^4}{r^3+g^3}
\end{equation}
is modified to eliminate the spacetime singularity.  Like the Kerr-Newman metric, the modification to Kerr is confined to a redefinition of $\Delta$, and thus upon choosing $N^2(r)$, $B^2(r)$, $F(r)$ and $f(r)$ as listed in \autoref{tab:altmetrics}, \autoref{eq:metric} reproduces the desired metric.

\subsection{Rotating Bardeen}
The rotating Bardeen metric is nearly identical to the rotating Hayward metric, with the exception that $\Delta_H$ is replaced with
\begin{equation}
    \Delta_{\rm B} = r^2 + a^2 - 2\frac{M r^4}{(r^2+g^2_*)^{3/2}},
\end{equation}
with the same attendant implications, summarized in \autoref{tab:altmetrics}.

\subsection{Kerr-Sen}
The Kerr-Sen metric describes a charged black hole that arises in a heterotic string theory.  It is characterized by an additional charge, related to a length scale via $r_{\rm S}=Q^2/4\pi\epsilon_0 M c^2$ \citep{Sen}.  In Boyer-Lindquist coordinates, this metric may be written as,
\begin{equation}
    \begin{split}
         ds^2 & = 
    -\left(1-\frac{2Mr}{\Sigma_{\rm S}}\right) dt^2
    - \frac{4Mra\sin^2\theta}{\Sigma_{\rm S}} dtd\phi\\ &
    +\left[r(r+r_{\rm S})+a^2+\frac{2Mra^2\sin^2\theta}{\Sigma_{\rm S}}\right]\sin^2\theta d\phi^2\\ &
    + \frac{\Sigma_{\rm S}}{\Delta_{\rm S}} dr^2
    + \Sigma_{\rm S} d\theta^2,
\label{eq:metric_sen}
    \end{split}
\end{equation}
where
\begin{equation}
\begin{aligned}
    \Sigma_{\rm S} &\equiv r(r+r_{\rm S}) + a^2\cos^2\theta\\
    \Delta_{\rm S} &\equiv r(r+r_{\rm S}) - 2Mr + a^2.
\end{aligned}
\end{equation}
Comparing the metric coefficients with those in \autoref{eq:metric}, it is immediately clear from the coefficient of $d\theta^2$ that $f(r)=r r_{\rm S}$.  Setting the remaining free functions,
\begin{equation}
\begin{aligned}
    F(r) &= r/[r(r+r_{\rm S}) + a^2]\\
    N^2(r) &= \Delta_{\rm S} F^2(r)\\
    B^2(r) &= r^2 F^2(r),
\end{aligned}
\end{equation}
brings \autoref{eq:metric} into agreement with \autoref{eq:metric_sen}.

An outer horizon occurs when
\begin{equation}
     r_+ = M - \frac{r_{\rm S}}{2} + \sqrt{ \left(m-\frac{r_{\rm S}}{2}\right)^2-a^2},
\end{equation}
which is well defined only for
\begin{equation}
    0\le r_{\rm S}\le2M
    ~~\text{and}~~
    |a| \le M-r_{\rm S}/2.
\end{equation}
The equation defining the polar photon sphere radius again reduces to a cubic polynomial,
\begin{equation}
    \begin{split}
        \rp^3 & - 3\left(M-\frac{r_{\rm S}}{2}\right) \rp^2 \\ &
    + \left(a^2-M r_{\rm S}+\frac{r_{\rm S}^2}{2}\right) \rp + a^2\left(M+\frac{r_{\rm S}}{2}\right) = 0,
    \end{split}
\end{equation}
which has solutions
\begin{equation}
    \begin{split}
        \rp & = M-\frac{r_{\rm S}}{2} + 2 M \sqrt{1-\frac{(a^2+2Mr_{\rm S}-r_{\rm S}^2/4)}{3M^2}} \\  & 
     \times \cos\left\{ \frac{1}{3} \cos^{-1}\left[ \frac{1-(a^2+Mr_{\rm S}-r_{\rm S}^2/4)/M^2}{\left[1-(a^2+2Mr_{\rm S}-r_{\rm S}^2/4)/3M\right]^{3/2}}\right]\right\}. 
    \end{split}
\end{equation}
which is again very similar to \autoref{eq:Kerr_rp}, with whcich the shadow size and Lyapunov exponent may be immediately computed.

\subsection{Rotating Simpson-Visser}
By applying Newman-Janis algorithm on the Simpson-Visser metric \citep{SimpsonVisser2019}, \citet{Shaikh2021} derives this metric in Boyer-Lindquist coordinates,
\begin{equation}
    \begin{split}
         ds^2 & = - \frac{\Delta}{\Sigma} (dt - a \sin^2\theta d\phi)^2\\ &
    + \frac{\sin^2\theta}{\Sigma} \left[ a dt - (r^2+a^2) d\phi \right]^2\\ &
    + \frac{\Sigma}{\Delta_{\rm SV}\Delta} dr^2 + \Sigma d\theta^2,
    \label{eq:simpson_visser}
    \end{split}
\end{equation}
where
\begin{equation}
    \Delta_{\rm SV}= 1-\frac{r_{\rm SV}^2}{r^2}.
\end{equation}
Here, $r_{\rm SV}$ is the radius of a non-singular minimal surface.  Comparing the metric coefficients with \autoref{eq:metric} we have:
\begin{equation}
\begin{aligned}
    f(r) &= 0\\
    F(r) &= u(r)\\
    N^2(r) &= \Delta u^2(r)\\
    B^2(r) &= \frac{r^2 u^2(r)}{\Delta_{\rm SV}}.
\end{aligned}
\end{equation}
See \autoref{tab:altmetrics} for a summery.

\subsection{Baines-Visser Metric}
In \citet{Visser2023} a family of metrics describing axisymmetric spacetimes is proposed that admits separable Hamilton-Jacobi and Klein-Gordon equations,
\begin{equation}
    \begin{split}
         ds^2 & = - \frac{\Delta_{\rm BV} e^{-2\Phi} - a^2\sin^2\theta}{\Xi^2+a^2\cos^2\theta} dt^2 \\ & 
    - 2 \frac{a\left(\Xi^2-\Delta_{\rm BV} e^{-2\Phi}+a^2\right)\sin^2\theta}{\Xi^2+a^2\cos^2\theta} dt d\phi\\ & 
    + \frac{\left[\left(\Xi^2+a^2\right)^2-e^{-2\Phi}\Delta_{\rm BV} a^2\sin^2\theta\right]\sin^2\theta}{\Xi^2+a^2\cos^2\theta} d\phi^2\\ & 
    + \frac{\Xi+a^2\cos^2\theta}{\Delta_{\rm BV}} dr^2
    + \left(\Xi^2+a^2\cos^2\theta\right) d\theta^2,
    \end{split}
\end{equation}
where $\Phi(r)$, $\Delta_{\rm BV}(r)$, and $\Phi(r)$ are three free functions of radius \citep[see Equation 2.3 of][]{Visser2023}.  Thus, this family metric is suggested as a useful class of foils for general relativity.  It is a special case of \autoref{eq:metric}, with 
\begin{equation}
\begin{aligned}
    f(r) &= \Xi(r) - (r^2+a^2)\\
    F(r) &= \frac{r}{\Xi(r)}\\
    N^2(r) &= \frac{r^2 \Delta_{\rm BV}(r) e^{-2\Phi(r)}}{ \Xi^2(r)} \\
    B^2(r) &= \frac{r^4 e^{-2\Phi(r)}}{\Xi^2(r)}.
\end{aligned}
\end{equation}
These are summarized in \autoref{tab:altmetrics}.

\subsection{Azreg-A\"inou Metric}
\label{app:AzregAinou}
\citet{AzregAinou2014} propose a family of metrics describing regular axisymmetric black hole spacetimes that possess a separable action with the following metric, described in Equation (25) of that paper (after changing the sign convention to match that in this paper),
\begin{equation}
    \begin{split}
         ds^2 &= -\left(1-\frac{2f_A}{\Sigma}\right) dt^2+\frac{2af_A\sin^2\theta}{\Sigma}dt d\phi \\ & 
    +\frac{[(r^2+a^2)^2-a^2\Delta_{\rm A}\sin^2\theta]\sin^2\theta}{\Sigma}d\phi^2\\ &
    +\frac{\Sigma}{\Delta_{\rm A}}dr^2+\Sigma d\theta^2,
    \label{eq:AAmetric}
    \end{split}
\end{equation}
where
\begin{equation}
   \Delta_{\rm A} = r^2+a^2-2f_A(r),
\end{equation}
and $f_A(r)$ is a free function of radius.  This metric is compatible with \autoref{eq:metric}, with 
\begin{equation}
\begin{aligned}
    f(r) &= 0\\
    F(r) &= u(r)\\
    N^2(r) &= \Delta_{\rm A} u^2(r) \\
    B^2(r) &= r^2 u^2(r).
\end{aligned}
\end{equation}
Again, these are summarized in \autoref{tab:altmetrics}.

\subsection{Rotating Ay\'on-Beato-Garcia Metric}
\citet{AzregAinou2014} present as a special case of their general family of regular axisymmetric black hole spacetimes (see \autoref{app:AzregAinou}) the rotating version of the metric proposed by \citet{AyonBeatoGarcia1998}, which describes a black hole in the presence of a nonlinear electrodynamic theory \citep{Salazar1987}.  This is obtained from \autoref{eq:AAmetric} by choosing,
\begin{equation}
    f_A(r) = \frac{2Mrr^4}{(r^2+Q^2)^{3/2}} - \frac{Q^2r^4}{(r^2+Q^2)^2},
\end{equation}
with the associated choices for $N^2(r)$, $B^2(r)$, $F(r)$, and $f(r)$ listed in \autoref{tab:altmetrics}.

\newpage
\bibliographystyle{aasjournal}
\bibliography{references.bib}

\end{document}